\documentclass[a4paper]{article}
\usepackage{amsmath}
\makeatletter
\def\cdash{--}

\def\bm#1{\mathchoice
	{\mbox{\boldmath$#1$}}%
          {\mbox{\boldmath$#1$}}%
          {\mbox{\boldmath$\scriptstyle#1$}}%
          {\mbox{\boldmath$\scriptscriptstyle#1$}}}%
\def\@refcitex[#1]#2{\if@filesw\immediate\write\@auxout 
	{\string\citation{#2}}\fi 
\def\@citea{}\@refcite{\@for\@citeb:=#2\do 
	{\@citea\def\@citea{, }\@ifundefined 
	{b@\@citeb}{{\bf ?}\@warning 
	{Citation `\@citeb' on page \thepage \space undefined}} 
	\hbox{\csname b@\@citeb\endcsname}}}{[#1]}} 
 \def\@refcite#1#2{{[#1]\if@tempswa\typeout 
        {WSPC warning: optional citation argument 
	ignored: `#2'} \fi}} 
 \def\refcite{\@ifnextchar[{\@tempswatrue
	\@refcitex}{\@tempswafalse\@refcitex[]}}
\makeatother

\def\hcentering#1{%
        \oddsidemargin=#1
        \advance\oddsidemargin-\textwidth%
        \oddsidemargin=0.5\oddsidemargin%
        \advance\oddsidemargin-1truein%
        \evensidemargin=\oddsidemargin}

\title{EXACTNESS IN THE PATH INTEGRAL OF THE COULOMB POTENTIAL IN ONE
SPACE DIMENSION}

\author{SEJII SAKODA\\
Department of Applied Physics,
National Defense Academy\\
Hashirimizu,
Yokosuka city,
Kanagawa 239-8686, Japan\\
\texttt{sakoda@nda.ac.jp}}
\begin{document}
\hcentering{210mm}
\maketitle

\begin{abstract}
We solve time-sliced path integrals of one-dimensional Coulomb
 system in an exact manner. In formulating path integrals, we make use
 of the Duru-Kleinert transformation with Fujikawa's gauge theoretical
 technique. Feynman kernels in the momentum representation both for
 bound states and scattering states will be obtained with clear pole
 structure that explains the exactness of the path integral. The
 path integrals presented here can be, therefore, evaluated exactly
 by making use of Cauchy's integral theorem.

\end{abstract}

\section{Introduction}

One-dimensional system with the Coulomb potential, or the hydrogen atom
in one space dimension, may find its application not so much in
practice.\footnote{After the completion of this work, the author has been
made aware of experimental studies on the exciton
spectra\cite{Loudon08} under the high magnetic field. It is shown in
Ref.\refcite{ElliottLoudon} that an exciton under extremely high
magnetic field can be well approximated by the one-dimensional Coulomb
system.} 
Nevertheless, after a work by Loudon\cite{Loudon}, a number of
publications on this apparently simple system have 
appeared\cite{Andrews66}\cdash\cite{SpecLee} to discuss the
peculiarity of the singular potential. 
The strange result obtained by Loudon was the twofold degeneracy in the
energy eigenvalues despite the one-dimensional nature of the
system. Another interesting but quite controversial issue was the
(non-)existence\cite{Andrews66,HainesRobert,Andrews76} of the ground
state of even parity with infinite binding energy. Exclusion of the even
states with finite energy\cite{GomeZim} and relativistic effect for the
removal of such unphysical states\cite{SpecLee} were also argued.
Among these works, it seems that only a few attempts on analyses from
the path integral formalism have so far been made. To our knowledge,
only in Ref.\refcite{Fischer95} descriptions of functional-integral
treatment, in  addition to the functional-analytic approach, of the
one-dimensional hydrogen atom can be found. The technique, based on the 
probabilistic measure, of the functional-integral employed in
Ref.\refcite{Fischer95} is rather mathematical than that we wish to
establish in this article. Furthermore the actual calculation of the
functional-integral is worked out by making use of some identity that
connects the Green function of the one-dimensional hydrogen atom to that
of a four-dimensional harmonic oscillator. In other words, the path
integral of the one-dimensional hydrogen atom is not still solved
explicitly. It cannot be, therefore, sufficient for our purpose to
understand the exactness of the path integral for one-dimensional
hydrogen atom by such calculation. 
As stated in Ref.\refcite{Fischer95}, there may exist some natural
prescription for the boundary condition of wave functions at the origin
required from the consistency of the path integral. We should be
able to find such structure by constructing path integrals in the
time-sliced form keeping connection with the operator formalism.
Therefore, from theoretical as well as pedagogical points of view, it
will be useful to develop another path integral technique for solving
such a singular system.

Path integral of the hydrogen atom was first successfully formulated by
Duru and Kleinert\cite{DK,DK2} by making use of the so called
Duru-Kleinert(DK) transformation. The DK transformation consists of two
fundamental ingredients: reparametrization of the time in the path
integral and making use of the ``square root of the coordinates'' by
means of a Kustaanheimo-Stiefel(KS) transformation\cite{KS}. Many
attempts, mainly on the use of KS transformation, by which we can write
the Coulomb path integral in terms of oscillator coordinates, with or
without time-slicing, have been made to clarify the approach by DK 
transformation\cite{RD}\cdash\cite{HK95}. 
It was Fujikawa\cite{KF} who first observed that the essence of the DK
transformation is the special choice of the gauge fixing condition for a
system with invariance under reparametrization of the time. He proposed
a new approach to formulate a Coulomb path integral in an exactly
solvable way from the view point of the Jacobi's principle of least
action. He also noticed that the use of KS transformation in the Coulomb
path integral is rather technical and demonstrated the exact solubility
of the Coulomb path integral in three dimension in terms of the
parabolic coordinates.

The hydrogen atom in one space dimension seems to be rather
exceptional in the path integral formalism. Although many examples with
detailed proofs are found in the textbook\cite{HK95} by Kleinert, the
space dimension $D_{\mathcal{C}}$ for the path integral of Coulomb
potential in chapter 14 of the Ref.\refcite{HK95} is
assumed to be $D_{\mathcal{C}}\ge2$. In particular, if we set
$D_{\mathcal{C}}=1$ in Eq.(14.93) of Ref.\refcite{HK95}, we obtain
$D_{\mathcal{O}}=0$ as the space dimension of the corresponding, or
DK-equivalent, oscillator coordinates. In fact, there does not seem to
be known any KS-type transformation for one-dimensional Coulomb system
excepting the one utilized in Ref.\refcite{Fischer95} as mentioned
above.  
Motivated by these observations, we adopt in this article Fujikawa's
formulation to find path integral descriptions for one-dimensional
Coulomb potential $V(x)=-\alpha/r$, where $r=\sqrt{x^{2}}$.   
Use of the KS transformation will be thus avoided in our formulation.
It was the problem of operator ordering, arising from nonlinear
canonical transformations such as KS transformation, in path integrals
that forced some authors to criticize the DK method with KS
transformation(see Ref.\refcite{HK87}, \refcite{HK95} and references
therein).
Our formulation of time-sliced path integrals in this article will be built 
basing on a symmetric Hamiltonian operator that has definite rule of
operator ordering, thereby being free from the ordering problem. The
effect of time-slicing from the formal continuum path integral will also
be problematic if we rely on such formal derivation of path
integrals. In our formalism, the basic ingredient is the short-time
kernel that describes an infinitesimal time evolution. 
Feynman kernels of finite time evolution will be constructed by
multiplication of such infinitesimal kernels. Without this process it
will be impossible to understand the mechanism of exactness in path
integrals considered here.

We organize this paper as follows:
in section 2 we will briefly explain Fujikawa's method then
formulate immediately a path integral with a negative fixed energy in
the Euclidean formalism. For the case of positive fixed energy, the
Euclidean technique does not work well. Hence it will be
treated separately for the usual Feynman kernel by setting the time
parameter to be real. For both of these path integrals we shall develop
a novel procedure in multiplication of infinitesimal kernels in the
momentum representation. The use of the momentum space wave functions is
due to the form of the symmetric Hamiltonian we set up. This will become 
clear soon in the next section.
Section 3 will  be devoted to the construction of the physical states
from the solution of path integrals in section 2. In
Fujikawa's procedure this is the important step for solving the problem
in terms of path integral. This will be achieved by analyses of the
resolvent operator. Conclusions and remarks will be found in the final
section.

\section{Path Integrals of the Free Particle with a Fixed Energy}	

For a system, whose Lagrangian being given by
$L_{\text{orig}}=\dot{x}^{2}/2-V(x)$,
we consider another Lagrangian with reparametrization invariance, given
by
\begin{equation}
\label{eq:KF001}
 L=\sqrt{2(E-V(x))\left(\frac{dx}{dt}\right)^{2}} 
\end{equation}
for a fixed energy $E$, according to the procedure introduced by
Fujikawa in Ref.~\refcite{KF}. The invariance of this new system under
a time reparametrization requires a gauge fixing upon quantization. As
was shown by Fujikawa, by choosing an appropriate gauge condition, we
can formulate a path integral that is equivalent to the DK transform of
the path integral for the original system.

The Lagrangian \eqref{eq:KF001} yields the canonical momentum, given by
\begin{equation}
 p=\sqrt{2(E-V(x))}\,
\frac{dx}{dt}\left/\sqrt{\left(\frac{dx}{dt}\right)^{2}}\right.,
\end{equation}
that results in a constraint
\begin{equation}
 \phi_{0}:=\frac{1}{2}p^{2}+V(x)-E=0.
\end{equation}
In addition, due to the fact $H=p\dot{x}-L=0$,
we need to test Dirac's total Hamiltonian, given by
\begin{equation}
 \label{eq:KF002}
H_{\text{T}}=u(x,p)\phi_{0}\approx0,
\end{equation}
in which the Lagrange multiplier $u(x,p)$ cannot be determined from
the equation of motion at all.
Choosing this rather arbitrary degree by hand exactly corresponds to the
choice of gauge fixing condition. The convenient way for us is setting
$u(x,p)=r$ to get
\begin{equation}
 \label{eq:KF003}
H_{\text{T}}=\frac{1}{2}rp^{2}-\alpha-Er,
\end{equation}
for the one-dimensional Coulomb potential. However, there is no
restriction in application of this procedure and gauge fixing to systems
other than the Coulomb system. We may, for example, enjoy this method
even for the free particle by setting $\alpha=0$ in the
above. Furthermore, if we solve path integrals of the free particle in
this formalism, the solution of corresponding ones for the Coulomb
system is immediate, since the difference in the Hamiltonian $H_{\text{T}}$ of
$\alpha\ne0$ from that of $\alpha=0$ is merely a constant. We therefore
first examine the quantization and its path integral of the total
Hamiltonian \eqref{eq:KF003} for the case of $\alpha=0$ in the following.

\subsection{Path integral for the negative energy}
Quantization of the system can be achieved by adopting a unitary
representation of the canonical commutation relation(CCR)
\begin{equation}
 \label{eq:ccr001}
[x,p]=i
\end{equation}
with the constraint \eqref{eq:KF002} as a physical state condition.
Since the Hamiltonian $H_{\text{T}}$ involves a term that consists of
$r$ multiplied by $p^{2}$, we need to specify an operator ordering upon
quantization. For brevity we choose a symmetric product for this term
to consider the operator Hamiltonian
\begin{equation}
 \label{eq:KF004}
\hat{H}_{\text{S}}=\frac{1}{2}\hat{p}\hat{r}\hat{p}-E\hat{r}.
\end{equation}
To elucidate the property of this Hamiltonian, we here put $E=-\omega^{2}/2$
and evaluate a Euclidean kernel(a Feynman kernel with imaginary time) in
the momentum representation:
\begin{equation}
 \label{eq:KF005}
\tilde{K}(p_{b},\tau_{b}|p_{a},\tau_{a})=
\langle{p_{b}}\vert e^{-\hat{H}_{\text{S}}(\tau_{b}-\tau_{a})}
\vert{p_{a}}\rangle,
\end{equation}
where $\vert{p}\rangle$ designates the momentum eigenvector. Since the
Hamiltonian \eqref{eq:KF004} is linear in $\hat{r}$, the integration
over $x$ in a path integral can be easily performed. This is the reason
why we take the momentum($p$) representation instead of the
coordinate($x$) representation.
 
The first step for the evaluation of the kernel is the calculation of an
infinitesimal kernel
\begin{equation}
 \label{eq:KF006}
\langle{p}\vert\left(1-\epsilon\hat{H}_{\text{S}}\right)
\vert{p{'}}\rangle=
\int_{-\infty}^{\infty}\!\frac{dx}{2\pi}\,
\exp\left[-i(p-p{'})x-\frac{r\epsilon}{2}\left(
pp{'}+\omega^{2}\right)\right],
\end{equation}
where $\epsilon=(\tau_{b}-\tau_{a})/N>0$ is taken to be $\omega\epsilon\ll1$.
We split the integration domain at $x=0$ and put $x\to-x$ for negative
$x$ to find
\begin{equation}
 \label{eq:KF0070}
\langle{p}\vert\left(1-\epsilon\hat{H}_{\text{S}}\right)
\vert{p{'}}\rangle
=
\sum_{\sigma=\pm1}
\int_{0}^{\infty}\!\frac{dx}{2\pi}\,
\exp\left[-\left\{
i(p-p{'})\sigma+\frac{\epsilon}{2}\left(
pp{'}+\omega^{2}\right)\right\}x\right].
\end{equation}
To ensure the convergence of this integration, a condition,
$pp{'}+\omega^{2}>0$, is necessary. Although it is not explicit, this
condition is equivalent to
$p^{2}+\omega^{2}>0$($p{'}{}^{2}+\omega^{2}>0$) that holds always for
real $p$($p{'}$). To check this we rewrite $pp{'}$ as $p^{2}-p(p-p{'})$
in the above to make a change of variable from $x$ to $(1+i\epsilon
p\sigma/2)x$. By taking the Jacobian into account, we find that the
right hand side of Eq.\eqref{eq:KF0070} is equivalent to
\begin{equation}
 \label{eq:KF0070b}
\sum_{\sigma=\pm1}
\int_{0}^{\infty}\!\frac{dx}{2\pi}\,
\exp\left[\frac{i\epsilon}{2}p\sigma-\left\{
i(p-p{'})\sigma+\frac{\epsilon}{2}\left(
p^{2}+\omega^{2}\right)\right\}x\right],
\end{equation}
up to $O(\epsilon)$ in the exponent.
Note that the additional term from the
Jacobian precisely matches to the one we need when rewriting
the symmetric product $\hat{p}\hat{r}\hat{p}$ as
$\hat{p}^{2}\hat{r}+[\hat{r},\hat{p}]$. 
The condition for convergence of the integration now reads
$p^{2}+\omega^{2}>0$ for this new expression. This proves the above
statement. We can therefore proceed to
carry out the integration with respect to $x$ to find
\begin{equation}
 \label{eq:KF007}
\langle{p}\vert\left(1-\epsilon\hat{H}_{\text{S}}\right)\vert{p{'}}\rangle=
\frac{1}{2\pi i}\,
\left[\frac{1}{
p-p{'}-i\epsilon\left(
pp{'}+\omega^{2}\right)/2
}-\frac{1}{
p-p{'}+i\epsilon\left(
pp{'}+\omega^{2}\right)/2
}
\right].
\end{equation}
Once we obtain the expression \eqref{eq:KF007}, we can regard it as a
function of two complex variables $p$ and $p{'}$. As a function of $p$,
each term in the right hand side has a simple pole at
\begin{equation}
 \label{eq:pole01}
p=\frac{p{'}+i\epsilon\omega^{2}/2}{1-i\epsilon p{'}/2},\
\frac{p{'}-i\epsilon\omega^{2}/2}{1+i\epsilon p{'}/2},
\end{equation}
respectively. As a confirmation, we observe that the real part of
 $pp{'}+\omega^{2}$ is positive for both of these poles if we set $p{'}$
 to be real in this evaluation. It will be interesting to see that the
 situation is quite similar to the one for an expression of a delta
 function
\begin{equation}
 \label{eq:KF008}
\delta(p-p^{'})=\frac{1}{2\pi i}\left(
\frac{1}{p-p{'}-i0_{+}}-\frac{1}{p-p{'}+i0_{+}}\right),
\end{equation}
where $0_{+}$ designates a positive infinitesimal. We may often utilize
this notation hereafter.

The second step for us to proceed is the multiplication of the
infinitesimal kernels. But before doing this we observe
\begin{equation}
 \label{eq:KF009}
p-p{'}\mp i\frac{\epsilon}{2}\left(pp{'}+\omega^{2}\right)=
(p-p{'})\cosh(\omega\epsilon/2)
\mp\frac{i}{\omega}\left(pp{'}+\omega^{2}\right)\sinh(\omega\epsilon/2)+
O(\epsilon^{2})
\end{equation}
so that we can imagine the kernel for a finite imaginary time
$\tau$($\tau=\tau_{b}-\tau_{a}$) to be
\begin{equation}
 \label{eq:KF010}
\tilde{K}(p_{b},\tau_{b}|p_{a},\tau_{a})\\
=
\sum_{\sigma=\pm1}
\frac{1}{2\pi i}
\frac{\sigma}{
(p_{b}-p_{a})\cosh(\omega\tau/2)-i\dfrac{\sigma}{\omega}
\left(p_{b}p_{a}+\omega^{2}\right)\sinh(\omega\tau/2)
}.
\end{equation}
Let us prove this by explicitly evaluating a product of infinitesimal
kernels. To this aim we recall the analogy to the delta function above
again. The first term of the infinitesimal kernel \eqref{eq:KF007}
possesses a simple pole of positive(negative) imaginary part as a
function of $p$($p{'}$) while the situation switches to the opposite in
the second term. We can therefore make use of the Cauchy's integral
formula, just like the case we will do in the multiplication of two
delta functions, in carrying out the integration
\begin{equation}
 \langle{p_{2}}\vert
\left(1-\epsilon\hat{H}_{\text{S}}\right)^{2}\vert{p_{0}}\rangle=
\int_{-\infty}^{\infty}\!dp_{1}\,
\langle{p_{2}}\vert\left(1-\epsilon\hat{H}_{\text{S}}\right)
\vert{p_{1}}\rangle\langle{p_{1}}\vert
\left(1-\epsilon\hat{H}_{\text{S}}\right)\vert{p_{0}}\rangle,
\end{equation}
to obtain
\begin{multline}
\label{eq:KF011}
\langle{p_{2}}\vert
\left(1-\epsilon\hat{H}_{\text{S}}\right)^{2}\vert{p_{0}}\rangle=
\frac{1}{2\pi i}
\times\left[\frac{1}{
(p_{2}-p_{0})\{1+(\omega\epsilon/2)^{2}\}-
i\omega\epsilon\dfrac{1}{\omega}\left(p_{2}p_{0}+\omega^{2}\right)
}\right.\\
\left.-\frac{1}{
(p_{2}-p_{0})\{1+(\omega\epsilon/2)^{2}\}+
i\omega\epsilon\dfrac{1}{\omega}\left(p_{2}p_{0}+\omega^{2}\right)
}
\right].
\end{multline}
This suggests us to suppose
\begin{equation}
\label{eq:KF012}
\begin{aligned}
& \langle{p_{j}}\vert
\left(1-\epsilon\hat{H}_{\text{S}}\right)^{j}\vert{p_{0}}\rangle=
\frac{1}{2\pi i}\\
&\times\left[\frac{1}{
(p_{j}-p_{0})C_{j}-
\dfrac{i}{\omega}S_{j}\left(p_{j}p_{0}+\omega^{2}\right)
}
-\frac{1}{
(p_{j}-p_{0})C_{j}+
\dfrac{i}{\omega}S_{j}\left(p_{j}p_{0}+\omega^{2}\right)
}
\right],
\end{aligned}
\end{equation}
with initial conditions, $C_{0}=1$ and $S_{0}=0_{+}$.
Then, by repeating the same procedure above, we
find that one more multiplication by a short-time kernel $\langle{p_{j+1}}\vert
\left(1-\epsilon\hat{H}_{\text{S}}\right)\vert{p_{j}}\rangle$ to this
expression results in 
\begin{equation}
\label{eq:KF013}
\begin{aligned}
& \langle{p_{j+1}}\vert
\left(1-\epsilon\hat{H}_{\text{S}}\right)^{j+1}\vert{p_{0}}\rangle=
\frac{1}{2\pi i}\\
&\times\sum_{\sigma=\pm1}
\frac{\sigma}{
(p_{j+1}-p_{0})(C_{j}+\omega\epsilon S_{j}/2)-
i\dfrac{\sigma}{\omega}(\omega\epsilon C_{j}/2+S_{j})
\left(p_{j+1}p_{0}+\omega^{2}\right)
}.
\end{aligned}
\end{equation}
By comparing Eq.\eqref{eq:KF013} with Eq.\eqref{eq:KF012}, we obtain a
set of recurrence formulas
\begin{equation}
 \label{eq:KF014}
C_{j+1}=C_{j}+\frac{\omega\epsilon}{2}S_{j},\quad
S_{j+1}=\frac{\omega\epsilon}{2}C_{j}+S_{j}.
\end{equation}
The solution of the above can be found immediately to be
\begin{equation}
 \label{eq:KF015}
C_{j}=\cosh(j\omega\epsilon/2),\quad
S_{j}=\sinh(j\omega\epsilon/2)
\end{equation}
to convince us that our conjecture in Eq.\eqref{eq:KF010} holds.

Eigenvalues and eigenfunctions of the Hamiltonian $\hat{H}_{\text{S}}$
can be extracted from the kernel in Eq.\eqref{eq:KF010}. To see this we
rewrite the first term of $\tilde{K}(p_{b},\tau_{b}|p_{a},\tau_{a})$ as
follows
\begin{equation}
 \label{eq:KF016}
\begin{aligned}
&\frac{1}{2\pi i}
\frac{1}{
(p_{b}-p_{a})\cosh(\omega\tau/2)-\dfrac{i}{\omega}\left(p_{b}p_{a}+\omega^{2}\right)\sinh(\omega\tau/2)
}\\
=&
\frac{\omega}{\pi}
\frac{1}{
e^{\omega\tau/2}(\omega+ip_{b})(\omega-ip_{a})-
e^{-\omega\tau/2}(\omega-ip_{b})(\omega+ip_{a})
}\\
=&
\frac{\omega}{\pi}
\frac{1}{(\omega+ip_{b})(\omega-ip_{a})}
\sum_{n=0}^{\infty}e^{-(n+1/2)\omega\tau}
\left(\frac{\omega-ip_{b}}{\omega+ip_{b}}\right)^{n}
\left(\frac{\omega+ip_{a}}{\omega-ip_{a}}\right)^{n}.
\end{aligned}
\end{equation}
In the same way, the second term is also expanded in a series as
\begin{equation}
 \label{eq:KF017}
\begin{aligned}
 &-\frac{1}{2\pi i}
\frac{1}{
(p_{b}-p_{a})\cosh(\omega\tau/2)+
\dfrac{i}{\omega}\left(p_{b}p_{a}+\omega^{2}\right)\sinh(\omega\tau/2)
}\\
=&
\frac{\omega}{\pi}
\frac{1}{(\omega-ip_{b})(\omega+ip_{a})}
\sum_{n=0}^{\infty}e^{-(n+1/2)\omega\tau}
\left(\frac{\omega+ip_{b}}{\omega-ip_{b}}\right)^{n}
\left(\frac{\omega-ip_{a}}{\omega+ip_{a}}\right)^{n}.
\end{aligned}
\end{equation}
We thus obtain another expression of the kernel in the series of
eigenfunctions, given by
\begin{equation}
 \label{eq:KF018}
\tilde{K}(p_{b},\tau_{b}|p_{a},\tau_{a})=
\sum_{n=0}^{\infty}e^{-(n+1/2)\omega\tau}
\left\{
\tilde{\psi}^{(+)}_{n}(p_{b})
\tilde{\psi}^{(+)}_{n}(p_{a}){}^{*}+
\tilde{\psi}^{(-)}_{n}(p_{b})
\tilde{\psi}^{(-)}_{n}(p_{a}){}^{*}
\right\},
\end{equation}
where $\tilde{\psi}^{(\pm)}_{n}(p)$ for $n=0,\,1,\,2,\,\dots$, defined as
\begin{equation}
 \label{eq:KF019}
\tilde{\psi}^{(\pm)}_{n}(p)=\sqrt{\frac{\omega}{\pi}}
\frac{1}{\omega\pm ip}\left(\frac{\omega\mp ip}{\omega\pm ip}\right)^{n},
\end{equation}
are eigenfunctions belonging to a common eigenvalue
$\lambda_{n}^{(\text{S})}(E)=(n+1/2)\omega$ in the momentum
representation. From these manipulation we observe the twofold
degeneracy in all eigenvalues of the Hamiltonian. Eigenfunctions in the
position representation can be obtained from
$\tilde{\psi}^{(\pm)}_{n}(p)$ as
\begin{equation}
 \label{eq:KF020}
\psi^{(\pm)}_{n}(x)=\int_{-\infty}^{\infty}\!\frac{dp}{\sqrt{2\pi}}\,
e^{ipx}\tilde{\psi}^{(\pm)}_{n}(p).
\end{equation}
Since $\tilde{\psi}^{(+)}_{n}(p)$($\tilde{\psi}^{(-)}_{n}(p)$) has a
pole of order $n+1$ at $p=i\omega$($p=-i\omega$), the eigenfunction
$\psi^{(+)}_{n}(x)$($\psi^{(-)}_{n}(x)$) has support only on
positive(negative) $x$. Hence we find
\begin{equation}
 \label{eq:KF021}
\begin{aligned}
\psi^{(+)}_{n}(x)=&\frac{(-1)^{n}}{n!}\theta(x)\sqrt{2\omega}
\left.\frac{d^{n}}{dp^{n}}\right\vert_{p=i\omega}
\left[
e^{ipx}(p+i\omega)^{n}
\right]\\
=&
(-1)^{n}\sqrt{2\omega}\theta(x)e^{-\omega x}L_{n}(2\omega x),
\end{aligned}
\end{equation}
where $L_{n}(z)$ designates a Laguerre polynomial of the $n$-th
order and $\theta(x)$ is the step function. In the same manner, by
calculating the residue at $p=-i\omega$, we obtain
\begin{equation}
 \label{eq:KF022}
\psi^{(-)}_{n}(x)=(-1)^{n}\sqrt{2\omega}\theta(-x)e^{\omega x}L_{n}(-2\omega x).
\end{equation}
In short, we may write
\begin{equation}
 \label{eq:KF023}
\psi^{(\pm)}_{n}(x)=(-1)^{n}\sqrt{2\omega}\theta(\pm x)e^{-\omega r}
L_{n}(2\omega r).
\end{equation}
In this way, by writing the kernel $\langle{x_{b}}\vert
e^{-\hat{H}_{\text{S}}\tau}\vert{x_{a}}\rangle$ as
$K(x_{b},\tau_{b}|x_{a},\tau_{a})$, we obtain
\begin{equation}
 \label{eq:KF024}
K(x_{b},\tau_{b}|x_{a},\tau_{a})=\sum_{n=0}^{\infty}e^{-(n+1/2)\omega\tau}
\left\{
\psi^{(+)}_{n}(x_{b})
\psi^{(+)}_{n}(x_{a}){}^{*}+
\psi^{(-)}_{n}(x_{b})
\psi^{(-)}_{n}(x_{a}){}^{*}
\right\},
\end{equation}
corresponding to Eq.~\eqref{eq:KF018}.

We have thus observed that a time-sliced path integral,
\begin{equation}
 \label{eq:KF025}
\int\prod_{i=0}^{N}\frac{dp_{i}}{2\pi}\,
\prod_{j=1}^{N}dx_{j}\,
\exp\left[
i\sum_{k=0}^{N}p_{k}(x_{k+1}-x_{k})-
\frac{\epsilon}{2}\sum_{k=1}^{N}\left(p_{k}p_{k-1}+\omega^{2}\right)r_{k}
\right]
\end{equation}
that describes a kernel $K(x_{b},\tau_{b}|x_{a},\tau_{a})$ in the limit
$N\to\infty$, can be evaluated exactly by making multiple use of
Cauchy's integral theorem. Therefore we can regard the model considered
here an interesting example of an exact path integral that can be
carried out without use of the Gaussian identity.

\subsection{Path integral for the positive energy}

So far our consideration has been restricted to the Hamiltonian
$\hat{H}_{\text{S}}=\hat{p}\hat{r}\hat{p}/2-E\hat{r}$ with a negative energy $E=-\omega^{2}/2$.
It will be also interesting to see how the path integral of this system
with a positive energy $E=k^{2}/2$($k>0$) can be solved. Hence our next
target to be examined here is the time-sliced path integral
\begin{equation}
 \label{eq:KF026}
\int\prod_{i=0}^{N}\frac{dp_{i}}{2\pi}\,
\prod_{j=1}^{N}dx_{j}\,
\exp\left[
i\sum_{j=0}^{N}p_{j}(x_{j+1}-x_{j})-
\frac{i\epsilon}{2}\sum_{j=1}^{N}\left(p_{j}p_{j-1}-k^{2}\right)r_{j}
\right].
\end{equation}
Unfortunately, however, we cannot define its Euclidean version without
restricting the range of values for $p$ as well as for the imaginary
time $\tau=-it$ due to the lack of positivity of the Hamiltonian in the
term proportional to $p^{2}-k^{2}$. We shall therefore briefly point out
some properties of the Euclidean path integral for this system at the end
of this subsection.

We begin again with evaluation of the kernel in the momentum representation.
By adding a positive infinitesimal to the imaginary part of $k^{2}$, we can
perform integration over $x$ in Eq.\eqref{eq:KF026} to obtain a short-time
kernel
\begin{equation}
 \label{eq:KF027}
\langle{p}\vert\left(1-i\epsilon\hat{H}_{\text{S}}\right)\vert{p{'}}\rangle\\
=
\frac{1}{2\pi i}\,
\left[\frac{1}{
(p-p{'})+\epsilon\left(
pp{'}-k^{2}\right)/2
}-\frac{1}{
(p-p{'})-\epsilon\left(
pp{'}-k^{2}\right)/2
}
\right].
\end{equation}
Thanks to the regularization factor, poles of each term in this kernel
are shifted off to the imaginary axis from the real line. For instance,
as a function of $p$, the kernel
$\langle{p}\vert\left(1-i\epsilon\hat{H}_{\text{S}}\right)\vert{p{'}}\rangle$
has poles at
\begin{equation}
 \label{eq:KF028}
p=\frac{p{'}+\epsilon k^{2}/2}{1+\epsilon p{'}/2}\quad
\text{and}\quad
\frac{p{'}-\epsilon k^{2}/2}{1-\epsilon p{'}/2},
\end{equation}
whose imaginary part being positive and negative, respectively.
Once we recognize this pole structure of the infinitesimal kernel, we
can immediately carry out integrations over $p$ for multiplication of
short-time kernels by making use of the similar technique we have
employed for the case of negative energy. We thus omit showing here the
detail of the procedure for finding a Feynman kernel
\begin{equation}
 \label{eq:KF029}
\tilde{K}(p_{b},t_{b}|p_{a},t_{a})\\
=
\sum_{\sigma=\pm1}
\frac{1}{2\pi i}
\frac{\sigma}{
(p_{b}-p_{a})\cosh(kt/2)+\dfrac{\sigma}{k}
\left(p_{b}p_{a}-k^{2}\right)\sinh(kt/2)
}
\end{equation}
in the momentum representation for a finite time $t=t_{b}-t_{a}>0$.
Note that the regulator for a short-time kernel is included even in this
expression as infinitesimal shifts $\mp i0_{+}$ for $\sigma=\pm1$ in the
denominator.

If we set $p=k\coth\theta$ or $p=k\tanh\theta$ corresponding to the sign
of $p^{2}-k^{2}$, it becomes easy to decompose the above kernel into an
integration over the continuous eigenvalue of $\hat{H}_{\text{S}}$. At
the outset, we set $p_{a}=k\tanh\theta_{a}$ and
$p_{b}=k\tanh\theta_{b}$. Then we rewrite each term in the kernel to
obtain
\begin{equation}
 \label{eq:KF030}
\frac{1}{(p_{b}-p_{a})\cosh(kt/2)\pm
\dfrac{1}{k}(p_{b}p_{a}-k^{2})\sinh(kt/2)}=
\frac{\cosh\theta_{b}\cosh\theta_{a}}
{k\sinh(\theta_{b}-\theta_{a}\mp kt/2)}.
\end{equation}
A useful formula for our aim here is an integral representation
\begin{equation}
\label{eq:KF031}
\frac{1}{\sinh x}=2i\int_{-\infty}^{\infty}\!dz\,
\frac{1}{1+e^{2\pi z}}
e^{2ixz},
\end{equation}
in which a condition, $-\pi<\Im(x)<0$, for the convergence being
necessary.
We here remember infinitesimal shifts $\mp i0_{+}$ in the Feynman kernel
\eqref{eq:KF029} to regard $\pm(\theta_{b}-\theta_{a})-kt/2$ as $x$ in
Eq.\eqref{eq:KF031}. Then by noticing that
$\sqrt{k^{2}-p^{2}}=k/\cosh\theta$ for $p=k\tanh\theta$, 
we can express the kernel Eq.\eqref{eq:KF029} as
\begin{equation}
 \label{eq:KF032}
\begin{aligned}
&\tilde{K}(p_{b},t_{b}|p_{a},t_{a})
=\frac{k}{\pi}
\frac{1}{\sqrt{(k^{2}-p_{b}^{2})(k^{2}-p_{a}^{2})}}\\
&\times
\sum_{\sigma=\pm1}
\int_{-\infty}^{\infty}\!d\nu\,
e^{-ikt\nu}
\frac{1}{1+e^{2\pi\nu}}
\left(\frac{k+p_{b}}{k-p_{b}}\right)^{i\sigma\nu}
\left(\frac{k+p_{a}}{k-p_{a}}\right)^{-i\sigma\nu}
\end{aligned}
\end{equation}
for a combination of $p_{a}$ and $p_{b}$ that satisfy
$\lvert{p_{a}}\rvert<k$ and $\lvert{p_{b}}\rvert<k$.

In the same way, for $p_{a}$ and $p_{b}$ satisfying
$\lvert{p_{a}}\rvert>k$, $\lvert{p_{b}}\rvert>k$ as well as
$p_{a}p_{b}-k^{2}>0$, we may set $p_{a}=k\coth\theta_{a}$ and
$p_{b}=k\coth\theta_{b}$ to find
\begin{equation}
 \label{eq:KF030A}
\frac{1}{(p_{b}-p_{a})\cosh(kt/2)\pm
\dfrac{1}{k}(p_{b}p_{a}-k^{2})\sinh(kt/2)}=
\frac{\sinh\theta_{b}\sinh\theta_{a}}
{-k\sinh(\theta_{b}-\theta_{a}\mp kt/2)},
\end{equation}
in which $\theta_{b}-\theta_{a}$ should be regarded as
$\theta_{b}-\theta_{a}\pm i0_{+}$ corresponding to the sign in front of
$kt/2$. We then utilize the formula \eqref{eq:KF031} again
to obtain
\begin{equation}
 \label{eq:KF032A}
\begin{aligned}
&\tilde{K}(p_{b},t_{b}|p_{a},t_{a})
=\frac{k}{\pi}
\frac{1}{\sqrt{(p_{b}^{2}-k^{2})(p_{a}^{2}-k^{2})}}\\
&\times
\sum_{\sigma=\pm1}
\int_{-\infty}^{\infty}\!d\nu\,
e^{-ikt\nu}
\frac{e^{2\pi\nu}}{1+e^{2\pi\nu}}
\left(\frac{p_{b}+k}{p_{b}-k}\right)^{i\sigma\nu}
\left(\frac{p_{a}+k}{p_{a}-k}\right)^{-i\sigma\nu}.
\end{aligned}
\end{equation}
for this case. Here we have made a change of
variable from $\nu$ to $-\nu$ and rewrite $1/(1+e^{-2\pi\nu})$ as
$e^{2\pi\nu}/(1+e^{2\pi\nu})$ to reach this expression.
Although there exist other possible combinations of values for $p_{a}$
and $p_{b}$, it will be enough to observe Eq.\eqref{eq:KF032} as well
as Eq.\eqref{eq:KF032A} for finding eigenvalues and eigenfunctions,
$\tilde{\psi}_{\lambda}^{(\pm)}(p)=\langle{p}\vert{\lambda;\pm}\rangle$,
of $\hat{H}_{\text{S}}$ in the momentum representation.
Explicitly, wave functions are given by
\begin{equation}
 \label{eq:KF033}
\tilde{\psi}_{\lambda}^{(\pm)}(p)=
\frac{1}{\sqrt{\pi(1+e^{2\pi\nu})}}
\frac{(k+p)^{(-1\pm 2i\nu)/2}}{(k-p)^{(1\pm 2i\nu)/2}}
\end{equation}
so that $\vert{\lambda;\pm}\rangle$ to be eigenvectors of
$\hat{H}_{\text{S}}$ with an eigenvalue $\lambda=k\nu$;
\begin{equation}
 \label{eq:KF034}
\hat{H}_{\text{S}}\vert{\lambda;\pm}\rangle=k\nu\vert{\lambda;\pm}\rangle.
\end{equation}
Note that the regularization for the short-time kernel
Eq.\eqref{eq:KF027} can now be interpreted as the prescription for
avoiding branch points on the complex $p$-plane. In the region
$\lvert{p}\rvert<k$, $p$ is shifted with an amount of $\mp i0_{+}$ in
$\tilde{\psi}_{\lambda}^{(\pm)}(p)$ by the effect of the regulator. On
the other hand, when $p$ is in regions where $\lvert{p}\rvert>k$, the
shifts occur in the opposite way. This specifies the way to go around
the branch points at $p=\pm k$. When $p$ moves along the real axis, a
detour that goes lower and upper side of $p=\pm k$ is made for
$\tilde{\psi}_{\lambda}^{(+)}(p)$ and for
$\tilde{\psi}_{\lambda}^{(-)}(p)$, respectively.
Due to these singularities of branch points, discontinuities
of wave functions as functions of a real variable $p$ at $p=\pm k$ are
unavoidable; if we consider the analytical continuation of the
eigenfunction $\psi_{\lambda}^{(+)}(p)$ in the region
$\lvert{p}\rvert<k$ to that of $p>k$, for instance, the wave function
gets a factor $e^{-i\pi/2+\pi\nu}$ multiplied in addition to a switch
from $k-p$ to $p-k$ in the denominator. Nevertheless, we can clarify the
completeness of the eigenfunctions simply by setting $t=0$ in each
expression, Eq.\eqref{eq:KF032}, \eqref{eq:KF032A}, and in other similar
ones, of the Feynman kernel. If we keep in mind the discontinuity
of eigenfunctions at branch points, we can also easily check the
following normalization and the orthogonality relations of
$\vert{\lambda;\pm}\rangle$,
\begin{equation}
 \label{eq:KF035}
\langle{\lambda;\sigma}\vert{\lambda{'};\sigma{'}}\rangle=
\delta_{\sigma,\sigma{'}}
\delta(\lambda-\lambda{'}).
\end{equation}
We have thus completed the calculation of the Feynman kernel in the
momentum representation.

Turning now to the original time-sliced path integral of
Eq.\eqref{eq:KF026} that defines a Feynman kernel in the position
diagonal representation, we make use of the integral representation of
the confluent hypergeometric, or Kummer, function to obtain
\begin{equation}
 \label{eq:KF036}
\frac{(k+p)^{(-1+2i\nu)/2}}{(k-p)^{(1+2i\nu)/2}}=
e^{\pi\nu}\int_{0}^{\infty}\!\!dx\,
e^{-i(p-k)x}F(\frac{1}{2}-i\nu,1,-2ikx),
\end{equation}
where the shift of $p$ by an amount of $-i0_{+}$ must be understood.
With the help of this formula, we can find the eigenfunction,
$\langle{x}\vert{\lambda;+}\rangle=\psi^{(+)}_{\lambda}(x)$, for an
eigenvalue $\lambda=k\nu$ in the position representation as
\begin{equation}
 \label{eq:KF037}
\psi^{(+)}_{\lambda}(x)=\theta(x)\sqrt{\frac{e^{\pi\nu}}{\cosh\pi\nu}}
e^{ikx}F(\frac{1}{2}-i\nu,1,-2ikx).
\end{equation}
In the same way, $\langle{p}\vert{\lambda;-}\rangle$ can be transformed
to $\langle{x}\vert{\lambda;-}\rangle=\psi^{(-)}_{\lambda}(x)$ as
\begin{equation}
 \label{eq:KF038}
\psi^{(-)}_{\lambda}(x)=\theta(-x)\sqrt{\frac{e^{\pi\nu}}{\cosh\pi\nu}}
e^{-ikx}F(\frac{1}{2}-i\nu,1,2ikx).
\end{equation}
These results enable us to transform the Feynman kernel
$\tilde{K}(p_{b},t_{b}|p_{a},t_{a})$ in the momentum representation to
the one in the position representation. Writing the Fourier transform of
$\tilde{K}(p_{b},t_{b}|p_{a},t_{a})$ as $K(x_{b},t_{b}|x_{a},t_{a})$, we
finally observe that the time-sliced path integral of
Eq.\eqref{eq:KF026} can be carried out exactly to yield
\begin{equation}
 \label{eq:KF039}
K(x_{b},t_{b}|x_{a},t_{a})
=
\int_{-\infty}^{\infty}\!d\lambda\,
e^{-i\lambda t}
\left\{
\psi_{\lambda}^{(+)}(x_{b})
\psi_{\lambda}^{(+)}(x_{a}){}^{*}+
\psi_{\lambda}^{(-)}(x_{b})
\psi_{\lambda}^{(-)}(x_{a}){}^{*}
\right\}.
\end{equation}
It will be obvious that the essence of the exact calculation in the
time-sliced path integral is again the pole structure of the short-time
kernel Eq.\eqref{eq:KF027} similar to the one we have found in
Eq.\eqref{eq:KF007} for the case of a negative energy $E=-\omega^{2}/2$.

As we have stated above at the beginning of this subsection, the
Euclidean path integral of the Hamiltonian $\hat{H}_{\text{S}}$ with a
positive energy $E=k^{2}/2$ cannot be freely defined. Nevertheless, we
may try to set $t=-i\tau$(Wick-rotation) in expressions,
Eq.\eqref{eq:KF032} and \eqref{eq:KF032A}, of the Feynman kernel
$\tilde{K}(p_{b},t_{b}|p_{a},t_{a})$ in the momentum representation. If
we make this change in Eq.\eqref{eq:KF032A}, we find a condition,
$0<k\tau<2\pi$, for the convergence of the integration over $\nu$. On
the other hand, Eq.\eqref{eq:KF032} can be Wick-rotated only for the
case $-2\pi<k\tau<0$ holds. Hence a kernel, though existed, for the
imaginary time $\tau$ cannot be extended its domain of definition into
other regions of momentum variables. For this reason we are restricted
to make considerations on Feynman kernels with $t$ in the real domain.

\section{Physical States}

We have clarified in the previous section that path integrals of the
Hamiltonian $\hat{H}_{\text{S}}$ for Feynman kernels with both positive
and negative energies can be performed in an exact manner. 
The Feynman kernel of $\hat{H}_{\text{S}}$ is, however, the one for the
unphysical system in a specific gauge. To get information of the
original system, we need to take the physical state condition
\eqref{eq:KF002} into account. This can be achieved by considering the
resolvent operator $(\hat{H}-E)^{-1}$, where $\hat{H}$ expresses the
Hamiltonian of the original system without gauge invariance. In this
section, we first examine the resolvent operator for a free particle
with a negative energy $E=-\omega^{2}/2$, then generalize the method to
fit the Coulomb system by making use of the result in the previous
section. Establishing the prescription to find the resolvent operator
from the Feynman kernel in these examples, we eventually solve the
scattering states of the one-dimensional Coulomb system. 

For the case of a free particle governed by the Hamiltonian
$\hat{H}=\hat{p}^{2}/2$, we first define non-symmetric operator
$\hat{H}_{\text{R}}=(\hat{H}-E)\hat{r}$($E=-\omega^{2}/2$) to observe
\begin{equation}
 \label{eq:RES01}
\left(\hat{H}-E\right)^{-1}=\hat{r}\hat{H}_{\text{R}}^{-1}.
\end{equation}
To evaluate the right hand side of the above in terms of the path
integral, we here invent a useful relation that connects
$\hat{H}_{\text{R}}$ to the symmetric one $\hat{H}_{\text{S}}$.
To this aim we introduce an operator $\hat{\sigma}$ by
$\hat{\sigma}=\hat{x}/\hat{r}$, which can be defined excepting the
origin $x=0$, to express the sign of the position operator.
Note that it is also possible to introduce $\hat{\sigma}$ by the
commutator $[\hat{r},\hat{p}]=i\hat{\sigma}$ as we have already done
implicitly in obtaining Eq.\eqref{eq:KF0070b}. As an important fact, it
should be remarked here that $\hat{\sigma}$ commutes with
$\hat{H}_{\text{S}}=\hat{p}\hat{r}\hat{p}/2-E\hat{r}$,
i.e. $\hat{\sigma}\hat{H}_{\text{S}}=\hat{H}_{\text{S}}\hat{\sigma}$.
It will be clear that the operator $\hat{\sigma}$ is responsible for the
supplementary quantum number $\sigma$ in the eigenfunctions of the
Hamiltonian $\hat{H}_{\text{S}}$. Needless to say, it is not a good
quantum number for Hamiltonians other than
$\hat{H}_{\text{S}}$. Nevertheless it will be useful for finding
eigenfunctions of the Coulomb Hamiltonian by restricting the domain of
$x$ to one of a half line corresponding to the value of $\sigma$. 
Keeping these in mind, we now define another operator
$\omega-i\hat{p}$ and its Hermitian conjugate to find
\begin{equation}
 \label{eq:RES02}
\hat{H}_{\text{R}}=\left(\omega\mp i\hat{p}\right)
\left(\hat{H}_{\text{S}}\pm\frac{\omega}{2}\hat{\sigma}\right)
\left(\omega\mp i\hat{p}\right)^{-1}.
\end{equation}
Though both of these formulas are equally useful for our purpose, we
here choose the one of $\omega-i\hat{p}$ for the formulation below.

By writing $\tilde{\psi}^{(\pm)}_{n}(p)$($n=0,\,1,\,2,\,\dots$) in
Eq.\eqref{eq:KF019} as $\langle{p}\vert{n;\pm}\rangle$, we denote right
eigenvectors of $\hat{H}_{\text{S}}$ as $\vert{n;\pm}\rangle$. Then we
define
$\vert\psi_{\text{R}}^{(n;\pm)}\rangle=(\omega-i\hat{p})\vert{n;\pm}\rangle$
to find
\begin{equation}
 \label{eq:RES03}
e^{-\hat{H}_{\text{R}}\tau}\vert{\psi_{\text{R}}^{(n;+)}}\rangle
=e^{-(n+1)\omega\tau}
\vert{\psi_{\text{R}}^{(n;+)}}\rangle,\quad
e^{-\hat{H}_{\text{R}}\tau}\vert{\psi_{\text{R}}^{(n;-)}}\rangle
=e^{-n\omega\tau}
\vert{\psi_{\text{R}}^{(n;-)}}\rangle.
\end{equation}
Completeness of the eigenvectors of $\hat{H}_{\text{S}}$ will
be expressed in terms of $\vert\psi_{\text{R}}^{(n;\pm)}\rangle$ and
their conjugates, defined by
$\langle\bar{\psi}_{\text{R}}^{(n;\pm)}\vert=\langle{n;\pm}\vert(\omega-i\hat{p})^{-1}$,
to yield
\begin{equation}
 \label{eq:RES04}
e^{-\hat{H}_{\text{R}}\tau}=
\sum_{n=0}^{\infty}\left\{
e^{-(n+1)\omega\tau}
\vert{\psi_{\text{R}}^{(n;+)}}\rangle
\langle{\bar{\psi}_{\text{R}}^{(n;+)}}\vert
+e^{-n\omega\tau}
\vert{\psi_{\text{R}}^{(n;-)}}\rangle
\langle{\bar{\psi}_{\text{R}}^{(n;-)}}\vert
\right\}.
\end{equation}
Sandwiching between $\langle{p_{b}}\vert$ and $\vert{p_{a}}\rangle$
this expression after multiplying $\hat{r}$ from the left, we obtain
\begin{equation}
 \label{eq:RES05}
\begin{aligned}
&\langle{p_{b}}\vert\hat{r}e^{-\hat{H}_{\text{R}}\tau}\vert{p_{a}}\rangle\\
=&
\sum_{n=0}^{\infty}\left\{
e^{-(n+1)\omega\tau}
\langle{p_{b}}\vert\hat{r}\vert{\psi_{\text{R}}^{(n;+)}}\rangle
\langle{\bar{\psi}_{\text{R}}^{(n;+)}}\vert{p_{a}}\rangle
+e^{-n\omega\tau}
\langle{p_{b}}\vert\hat{r}\vert{\psi_{\text{R}}^{(n;-)}}\rangle
\langle{\bar{\psi}_{\text{R}}^{(n;-)}}\vert{p_{a}}\rangle
\right\}.
\end{aligned}
\end{equation}
Evaluation of the operator $\hat{r}$ in the above is done as follows:
we may write $\hat{r}=\hat{x}\hat{\sigma}$ in which $\hat{x}$ has clear
definition in the momentum representation, given by
$\langle{p}\vert\hat{x}=i\partial\langle{p}\vert/\partial p$, our task
reduces to calculate
$\langle{p}\vert\hat{\sigma}\vert{\psi_{\text{R}}^{(n;\sigma)}}\rangle$
for $\sigma=\pm1$. By noticing
$[\hat{\sigma},\hat{p}]=2i\delta(\hat{x})$, where the delta function of
$\hat{x}$ being defined by
\begin{equation}
\label{eq:RES06}
\delta(\hat{x})=\frac{1}{2\pi}\int_{-\infty}^{\infty}\!\!dp{'}\,
e^{ip{'}\hat{x}}
\end{equation}
so that
\begin{equation}
\label{eq:RES07}
\langle{p}\vert\delta(\hat{x})=
\frac{1}{2\pi}\int_{-\infty}^{\infty}\!\!dp{'}\,
\langle{p{'}}\vert,\quad
\langle{p{'}}\vert\hat{p}=p{'}\langle{p{'}}\vert,
\end{equation}
we rewrite
$\langle{p}\vert\hat{\sigma}\vert{\psi_{\text{R}}^{(n;\sigma)}}\rangle$
as
\begin{equation}
 \label{eq:RES08}
\langle{p}\vert\left\{(\omega-i\hat{p})\hat{\sigma}-
i[\hat{\sigma},\hat{p}]\right\}\vert{n;\sigma}\rangle=
\sigma(\omega-ip)\langle{p}\vert{n;\sigma}\rangle+
\frac{1}{\pi}\int_{-\infty}^{\infty}\!\!dp{'}\,
\langle{p{'}}\vert{n;\sigma}\rangle.
\end{equation}
Since the second term above is independent of $p$ to be destroyed by
differentiation with respect to $p$, only the first term contributes to 
$\langle{p}\vert\hat{r}\vert{\psi_{\text{R}}^{(n;\sigma)}}\rangle$
to yield
\begin{equation}
 \label{eq:RES09}
\langle{p}\vert\hat{r}\vert{\psi_{\text{R}}^{(n;\pm)}}\rangle=
\pm i\frac{\partial}{\partial p}\left\{
(\omega-ip)\langle{p}\vert{n;\pm}\rangle
\right\}.
\end{equation}
Remembering the definition of wave functions
$\tilde{\psi}_{n}^{(\pm)}(p)=\langle{p}\vert{n;\pm}\rangle$ in
Eq.\eqref{eq:KF019}, we may write explicitly as
\begin{equation}
 \label{eq:RES10}
\begin{aligned}
\langle{p}\vert\hat{r}\vert{\psi_{\text{R}}^{(n;+)}}\rangle=&
\sqrt{\frac{\omega}{\pi}}\frac{2(n+1)\omega}{\omega^{2}+p^{2}}
\left(\frac{\omega-ip}{\omega+ip}\right)^{n+1},\\
\langle{p}\vert\hat{r}\vert{\psi_{\text{R}}^{(n;-)}}\rangle=&
\sqrt{\frac{\omega}{\pi}}\frac{2n\omega}{\omega^{2}+p^{2}}
\left(\frac{\omega+ip}{\omega-ip}\right)^{n}.
\end{aligned}
\end{equation}
Here it must be remarked that
$\langle{p}\vert\hat{r}\vert{\psi_{\text{R}}^{(n;-)}}\rangle$ vanishes
for $n=0$ thereby disappearing from the series in Eq.\eqref{eq:RES05}.
Combining the result above with those of
$\langle{\bar{\psi}_{\text{R}}^{(n;\pm)}}\vert{p_{a}}\rangle$, we find
\begin{equation}
 \label{eq:RES11}
\begin{aligned}
\langle{p_{b}}\vert\hat{r}e^{-\hat{H}_{\text{R}}\tau}\vert{p_{a}}\rangle
=&
\sum_{n=1}^{\infty}e^{-n\omega\tau}
\frac{\omega}{\pi}\frac{2n\omega}
{(\omega^{2}+p_{b}^{2})(\omega^{2}+p_{a}^{2})}\\
&\times
\left\{
\left(\frac{\omega-ip_{b}}{\omega+ip_{b}}
\frac{\omega+ip_{a}}{\omega-ip_{a}}\right)^{n}
+\left(\frac{\omega+ip_{b}}{\omega-ip_{b}}
\frac{\omega-ip_{a}}{\omega+ip_{a}}\right)^{n}
\right\}.
\end{aligned}
\end{equation}
We are now at the point to observe, by integrating the above with
respect to $\tau$, that
\begin{equation}
 \label{eq:RES12}
\langle{p_{b}}\vert\frac{1}{\hat{H}-E}\vert{p_{a}}\rangle
=
\frac{2}
{(\omega^{2}+p_{b}^{2})}\delta(p_{b}-p_{a})-
\frac{\omega}{\pi}
\frac{2}
{(\omega^{2}+p_{b}^{2})(\omega^{2}+p_{a}^{2})}
\end{equation}
for $E=-\omega^{2}/2$. It is precisely the first term of the right hand
side that should be defined by the left hand side.
Therefore the second term above
indicates a contradiction in our manipulation. The origin of this
obvious disagreement is the disappearance of
$\langle{p}\vert\hat{r}\vert{\psi_{\text{R}}^{(0;-)}}\rangle$ from the
sum. A quick recovery for this discrepancy will be possible if we keep
$\langle{p}\vert\hat{r}\vert{\psi_{\text{R}}^{(0;-)}}\rangle$ until we
carry out the integration over $\tau$, by replacing it with
$e^{-\nu\omega\tau}\langle{p}\vert\hat{r}\vert{\psi_{\text{R}}^{(\nu;-)}}\rangle$,
thereupon letting $\nu\to0$. To
be consistent, however, we should examine this phenomenon in detail
otherwise we will not be successful in establishing the method to define
the operator $\hat{r}\hat{H}_{\text{R}}^{-1}$. Though interesting, it
will not be so useful to look more closely at the resolvent operator
with a negative energy for the free particle. Hence we terminate the
analysis on this system here by learning that the definition of the
resolvent operator by its action on each eigenfunction of the
non-symmetric operator $\hat{H}_{\text{R}}$ fails due to the
disappearance of an eigenfunction from the complete set.

We now switch the Hamiltonian to that of the Coulomb system by changing
$\hat{H}_{\text{R}}$ to $\hat{H}_{\text{R}}-\alpha$ in the above
argument.
Observation of the failure in the naive approach above by integration
over $\tau$ term by term on each eigenfunction of $\hat{H}_{\text{R}}$
suggests us to try dealing with each term in the Euclidean kernel
\begin{equation}
 \label{eq:RES13}
\begin{aligned}
&\langle{p_{b}}\vert
e^{-(\hat{H}_{\text{S}}-\alpha)\tau}
\vert{p_{a}}\rangle\\
=&
\sum_{\sigma=\pm1}
\frac{1}{2\pi i}
\frac{\sigma}{
(p_{b}-p_{a})\cosh(\omega\tau/2)-i\dfrac{\sigma}{\omega}
\left(p_{b}p_{a}+\omega^{2}\right)\sinh(\omega\tau/2)
} e^{\alpha\tau}
\end{aligned}
\end{equation}
as a whole. To this aim, let us set $p_{b}=\omega\tan\theta_{b}$ and
$p_{a}=\omega\tan\theta_{a}$ then rewrite the above in terms of these
new variables.
Firstly, we write
\begin{equation}
\label{eq:RES014}
\begin{aligned}
 &(p_{b}-p_{a})\cosh(\omega\tau/2)\mp\frac{i}{\omega}
\left(p_{b}p_{a}+\omega^{2}\right)\sinh(\omega\tau/2)\\
=&
\frac{\omega}{\cos\theta_{b}\cos\theta_{a}}
\sin(\theta_{b}-\theta_{a}\mp i\omega\tau/2)
\end{aligned}
\end{equation}
to find
\begin{equation}
\label{eq:RES14}
\begin{aligned}
&\langle{p_{b}}\vert\hat{r}
e^{-(\hat{H}_{\text{R}}-\alpha)\tau}
\vert{p_{a}}\rangle\\
=&
\sum_{\sigma=\pm1}
\frac{\cos^{2}\theta_{b}\cos^{2}\theta_{a}}{2\pi\omega^{2}}
\frac{\partial}{\partial\theta_{b}}
\frac{e^{-i(\theta_{b}-\theta_{a}-i\sigma\omega\tau/2)+\alpha\tau}}
{\sin(\theta_{b}-\theta_{a}-i\sigma\omega\tau/2)}\\
=& 
-\frac{\cos^{2}\theta_{b}\cos^{2}\theta_{a}}{2\pi\omega^{2}}
\left[\frac{e^{\alpha\tau}}
{\sin^{2}(\theta_{b}-\theta_{a}-i\omega\tau/2)}
+
\frac{e^{\alpha\tau}}
{\sin^{2}(\theta_{b}-\theta_{a}+i\omega\tau/2)}
\right].
\end{aligned}
\end{equation}
Since we are dealing with the case $\alpha>0$(attractive Coulomb
potential), both of the terms in the above monotonically tends
to $0$ if we let $\tau\to-\infty$. Although the path integral for the
Euclidean kernel has been performed only for $\tau>0$, the result can be
regarded as a function of a complex variable $\tau$, besides being a
function of $p_{b}$ and $p_{a}$. The above mentioned behavior in the
asymptotic region, $\tau\to-\infty$, naturally suggests us to integrate
Eq.\eqref{eq:RES14} with respect to $\tau$ form $0$ to $-\infty$ for a
definition of $\hat{r}(\hat{H}_{\text{R}}-\alpha)^{-1}$. However, when
the interaction is turned off by setting $\alpha=0$, the behavior of
the first term becomes wrong. This will be confirmed easily by looking
at the second line in the above for $\sigma=+1$ and
$\tau\to-\infty$. Such a discrepancy must be avoided for the consistency
of the formalism. We therefore choose the domain for integration over
$\tau$ from $0$ to $+\infty$ for $\sigma=+1$ and from $0$ to $-\infty$
for $\sigma=-1$ with the assumption $\omega>\alpha$ for a while.
Note that, by this new prescription, we can explain the inconsistency
found in Eq.\eqref{eq:RES12} above. The undesirable term is the
consequence of the non-vanishing, though finite, behavior of the second
term in Eq.\eqref{eq:RES14} when integrated over $\tau$ from $0$ to
$+\infty$($\alpha=0$ should be set for the free particle).

We here utilize the integral representation for a hypergeometric
function to carry out the integration of Eq.\eqref{eq:RES14} over
$\tau$, thereby obtaining
\begin{equation}
 \label{eq:RES15}
\int_{0}^{\infty}\!\!\!
\frac{e^{\pm\alpha\tau}\,d\tau}
{\sin^{2}(\theta_{b}-\theta_{a}-i\omega\tau/2)}=
-\frac{4}{\omega}\frac{e^{-2i(\theta_{b}-\theta_{a})}}{1\mp\hat{\alpha}}
F(2,1\mp\hat{\alpha},2\mp\hat{\alpha};e^{-2i(\theta_{b}-\theta_{a})}),
\end{equation}
where $\hat{\alpha}=\alpha/\omega$ and $F(a,b,c;z)$ denotes a
hypergeometric function. Here we should recall the initial condition,
$S_{0}=0_{+}$, imposed in the derivation of the Euclidean kernel. We
have chosen $0_{+}$ instead of $0$ to make the kernel reproduce the delta
function $\delta(p_{b}-p_{a})$ even after taking the limit
$\tau\to0$. This prescription now takes effect in the above by
multiplying $e^{\mp0_{+}}$ to $e^{-2i(\theta_{b}-\theta_{a})}$
depending on the sign in front of $\hat{\alpha}$.
In view of this prescription, we should rewrite one of the above as
\begin{equation}
 \label{eq:RES015}
-\frac{z}{1+\hat{\alpha}}
F(2,1+\hat{\alpha},2+\hat{\alpha};z)\\
=
\frac{1}{1-\hat{\alpha}}\frac{1}{z}
F(2,1-\hat{\alpha},2-\hat{\alpha};\frac{1}{z})
+\frac{\pi\hat{\alpha}}{\sin\pi\hat{\alpha}}
(-z)^{-\hat{\alpha}}
\end{equation}
for $z=e^{-2i(\theta_{b}-\theta_{a}+i0_{+})}$ to be useful as a
hypergeometric series. Taking these into account, we obtain
\begin{equation}
 \label{eq:RES16}
\begin{aligned}
&\langle{p_{b}}\vert
\frac{1}{\hat{H}_{\text{C}}-E-i0_{+}}
\vert{p_{a}}\rangle
=
\frac{\omega}{\pi}
\frac{2}{(\omega^{2}+p_{b}^{2})(\omega^{2}+p_{a}^{2})}\\
&\times
\left\{\frac{1}{1-\hat{\alpha}}\left(
e^{-2i(\theta_{b}-\theta_{a}-i0_{+})}
F(2,1-\hat{\alpha},2-\hat{\alpha};e^{-2i(\theta_{b}-\theta_{a}-i0_{+})})
\right.\right.\\
&\left.\left.+
e^{2i(\theta_{b}-\theta_{a}+i0_{+})}
F(2,1-\hat{\alpha},2-\hat{\alpha};e^{2i(\theta_{b}-\theta_{a}+i0_{+})})
\right)\right.\\
&\left.+
\frac{\pi\hat{\alpha}}{\sin\pi\hat{\alpha}}
\left(-e^{-2i(\theta_{b}-\theta_{a}+i0_{+})}\right)^{-\hat{\alpha}}
\right\}
\end{aligned}
\end{equation}
for the Hamiltonian $\hat{H}_{\text{C}}=\hat{p}^{2}/2-\alpha/\hat{r}$
with $E=-\omega^{2}/2$. Here use has been made of the fact that the
effect of our prescription of adding $\pm0_{+}$ in the denominator of
the kernel is equivalent to regarding $E=-\omega^{2}/2$ as to be
associated with $+i0_{+}$. Series expansions of hypergeometric
functions in the above will then bring us
\begin{equation}
 \label{eq:RES17}
\begin{aligned}
&\langle{p_{b}}\vert
\frac{1}{\hat{H}_{\text{C}}-E-i0_{+}}
\vert{p_{a}}\rangle
=
\frac{\omega}{\pi}
\frac{2}{(\omega^{2}+p_{b}^{2})(\omega^{2}+p_{a}^{2})}\\
&\times
\left[\sum_{n=1}^{\infty}\frac{n}{n-\hat{\alpha}}
\left\{
\left(\frac{\omega-ip_{b}}{\omega+ip_{b}}
\frac{\omega+ip_{a}}{\omega-ip_{a}}
\right)^{n}+
\left(\frac{\omega+ip_{b}}{\omega-ip_{b}}
\frac{\omega-ip_{a}}{\omega+ip_{a}}
\right)^{n}
\right\}
\right.\\
&\left.+
\frac{2\pi i\hat{\alpha}}{1-e^{-2\pi i\hat{\alpha}}}
\left(\frac{\omega-ip_{b}}{\omega+ip_{b}}
\frac{\omega+ip_{a}}{\omega-ip_{a}}\right)^{-\hat{\alpha}}
\right].
\end{aligned}
\end{equation}
Clearly, there exist poles of the Green function \eqref{eq:RES17} at
$E=E_{n}$, 
\begin{equation}
 \label{eq:RES18}
E_{n}=-\frac{\omega_{n}^{2}}{2},\
\omega_{n}=\frac{\alpha}{n},\
\end{equation}
for $n=1,\,2,\,3,\dots$, corresponding to bound states with twofold
degeneracy for each energy eigenvalue.
By approximating $\omega(E)=\sqrt{-2E}$ as
$\omega(E)\simeq\omega_{n}+(E_{n}-E)/\omega_{n}$ in the vicinity of
$n$-th eigenvalue, we obtain
\begin{equation}
 \label{eq:RES19}
\begin{aligned}
&\frac{\omega}{\pi}
\frac{2}{(\omega^{2}+p_{b}^{2})(\omega^{2}+p_{a}^{2})}
\frac{n}{n-\hat{\alpha}}
\left(\frac{\omega\mp ip_{b}}{\omega\pm ip_{b}}
\frac{\omega\pm ip_{a}}{\omega\mp ip_{a}}
\right)^{n}\\
\simeq &
\frac{1}{E_{n}-E}
\frac{2}{\pi}\left(\frac{\alpha}{n}\right)^{3}
\frac{1}{(\omega_{n}^{2}+p_{b}^{2})(\omega_{n}^{2}+p_{a}^{2})}
\left(\frac{\omega_{n}\mp ip_{b}}{\omega_{n}\pm ip_{b}}
\frac{\omega_{n}\pm ip_{a}}{\omega_{n}\mp ip_{a}}
\right)^{n}
\end{aligned}
\end{equation}
to extract the pole parts of the Green function as
\begin{equation}
 \label{eq:RES20}
\langle{p_{b}}\vert
\frac{1}{\hat{H}_{\text{C}}-E-i0_{+}}
\vert{p_{a}}\rangle
\simeq
\sum_{n=1}^{\infty}\frac{1}{E_{n}-E}
\left\{
\tilde{\psi}_{\text{C},n}^{(+)}(p_{b})
\tilde{\psi}_{\text{C},n}^{(+)}(p_{a}){}^{*}
+
\tilde{\psi}_{\text{C},n}^{(-)}(p_{b})
\tilde{\psi}_{\text{C},n}^{(-)}(p_{a}){}^{*}
\right\}
\end{equation}
where
\begin{equation}
\label{eq:RES21}
\tilde{\psi}_{\text{C},n}^{(\pm)}(p)
=
\sqrt{\frac{2}{\pi}}\left(\frac{\alpha}{n}\right)^{3/2}
\frac{1}{\omega_{n}^{2}+p^{2}}
\left(\frac{\omega_{n}\mp ip}{\omega_{n}\pm ip}
\right)^{n}.
\end{equation}
We have thus clarified the pole structure of the Green function.
Apart from the continuous spectrum, poles of the Green function
precisely describe the zeros of $\hat{H}_{\text{C}}-E$. The gauge fixing
condition, given by $H_{\text{T}}\simeq0$, in our formulation of the
corresponding classical system is proportional to $H_{\text{C}}-E$ with
an almost non-vanishing factor $r$. Therefore we may regard enforcing
$\hat{H}_{\text{C}}-E=0$ in quantum mechanics as the implementation of the
gauge fixing condition that picks up physical or admissible states from
the Hilbert space of the system with a gauge degree.

Another Green function that reflects different boundary condition is
obtained by changing the prescription from $E\to E+i0_{+}$ to 
$E\to E-i0_{+}$, instead. For this case, poles appear at
$\omega=-\omega_{n}$. These cannot be regarded as bound states
since $\Re(\omega)\ge0$ is needed in setting
$E=-\omega^{2}/2$. Wave functions for continuous eigenvalues can be
found by subtracting $(\hat{H}_{\text{C}}-E+i0_{+})^{-1}$ from
$(\hat{H}_{\text{C}}-E-i0_{+})^{-1}$. It will be useful here to consider
in the region $E=k^{2}/2\ge0$ to find the continuous eigenfunctions.
The transition from a negative energy to a positive one can be most
easily achieved by setting $\omega\to-ik$,
$\theta_{b,a}\to i\theta_{b,a}$ as well as writing $\alpha/k$ as
$\hat{\alpha}$ again. The expression for $E=-\omega^{2}/2$ in
Eq.\eqref{eq:RES14} then becomes
\begin{equation}
\label{eq:RES22}
\begin{aligned}
&\langle{p_{b}}\vert\hat{r}
e^{-i(\hat{H}_{\text{R}}-\alpha)t}
\vert{p_{a}}\rangle\\
=& 
-\frac{\cosh^{2}\theta_{b}\cosh^{2}\theta_{a}}{2\pi k^{2}}
\left[\frac{e^{i\alpha t}}
{\sinh^{2}(\theta_{b}-\theta_{a}-ikt/2)}
+
\frac{e^{i\alpha t}}
{\sinh^{2}(\theta_{b}-\theta_{a}+ikt/2)}
\right].
\end{aligned}
\end{equation}
This expression will be integrated over $t$ from $0$ to
$+\infty$($-\infty$) for the first(second) term to yield
\begin{equation}
 \label{eq:RES23}
\begin{aligned}
&\langle{p_{b}}\vert
\frac{1}{\hat{H}_{\text{C}}-E-i0_{+}}
\vert{p_{a}}\rangle
=
\frac{k}{\pi}
\frac{2}{(k^{2}-p_{b}^{2})(k^{2}-p_{a}^{2})}\\
&\times
\int_{-\infty}^{\infty}\!\!\frac{d\nu}{e^{2\pi\nu}-1}
\frac{\nu}{\nu-\hat{\alpha}}
\left\{
\left(\frac{k+p_{b}}{k-p_{b}}
\frac{k-p_{a}}{k+p_{a}}
\right)^{i\nu}+
\left(\frac{k-p_{b}}{k+p_{b}}
\frac{k+p_{a}}{k-p_{a}}
\right)^{i\nu}
\right\},
\end{aligned}
\end{equation}
in which $\hat{\alpha}=\alpha/k$ has a negative imaginary part due to
the boundary condition of the Green function: 
$k\to k+i0_{+}$ for $E\to E+i0_{+}$. Here we have made use of the
formula Eq.\eqref{eq:KF031} again in obtaining the above.
If we change the boundary condition by replacing this with 
$E\to E-i0_{+}$, the sign of the imaginary part of $\hat{\alpha}$ is
also changed as well. Then, by subtraction of these two Green functions,
we get a delta function of $\nu-\hat{\alpha}$ to find
\begin{equation}
 \label{eq:RES24}
\begin{aligned}
&\frac{1}{2\pi i}\langle{p_{b}}\vert
\left[
\frac{1}{\hat{H}_{\text{C}}-E-i0_{+}}
-\frac{1}{\hat{H}_{\text{C}}-E+i0_{+}}
\right]
\vert{p_{a}}
\rangle
=
\frac{k}{\pi}
\frac{2}{(k^{2}-p_{b}^{2})(k^{2}-p_{a}^{2})}\\
&\times
\frac{\hat{\alpha}}{e^{2\pi\hat{\alpha}}-1}
\left\{
\left(\frac{k+p_{b}}{k-p_{b}}
\frac{k-p_{a}}{k+p_{a}}
\right)^{i\hat{\alpha}}+
\left(\frac{k-p_{b}}{k+p_{b}}
\frac{k+p_{a}}{k-p_{a}}
\right)^{i\hat{\alpha}}
\right\}
\end{aligned}
\end{equation}
by omitting the pole parts. From this expression, eigenfunctions for
$E=k^{2}/2$ immediately read
\begin{equation}
 \label{eq:RES25}
\tilde{\psi}_{\text{C},k}^{(\pm)}(p)=
\sqrt{\frac{\alpha e^{-\pi\hat{\alpha}}}{\pi\sinh\pi\hat{\alpha}}}
\frac{1}{k^{2}-p^{2}}
\left(\frac{k\pm p}{k\mp p}
\right)^{i\hat{\alpha}}.
\end{equation} 
Corresponding eigenfunctions in coordinate representation are also
obtained by Fourier transformation as
\begin{equation}
 \label{eq:RES26}
\psi_{\text{C},k}^{(\pm)}(x)=2\sqrt{\frac{\pi\alpha e^{\pi\hat{\alpha}}}
{\sinh\pi\hat{\alpha}}}
\theta(\pm x)e^{ikr}rF(1-i\hat{\alpha},2,-2ikr),\ r=\lvert x\rvert.
\end{equation}
In the same way, eigenfunctions for bound states, given by
Eq.\eqref{eq:RES21}, can be transformed to those of coordinate
representation to yield
\begin{equation}
 \label{eq:RES27}
\psi_{\text{C},n}^{(\pm)}(x)=2(-1)^{n+1}\omega_{n}^{3/2}
\theta(\pm x)e^{-\omega_{n}r}rF(1-n,2,2\omega_{n}r),\
\omega_{n}=\frac{\alpha}{n},\ r=\lvert x\rvert.
\end{equation}
Since $\sigma$ is not suitable for description of the Coulomb system, we
may consider linear combinations of wave functions above for
construction of eigenfunctions with even and odd parities. This completes
the process of solving the Coulomb system in terms of path integral.

To close this section let us add a comment on the difference in the
number of bound states of the Coulomb system from that of
$H_{\text{S}}$. We have observed that, through the procedure of
constructing $(\hat{H}_{\text{C}}-E)^{-1}$ from the eigenfunctions of
$\hat{H}_{\text{S}}$ with the aid of $\omega-i\hat{p}$ and its inverse,
one of the eigenvectors of $\hat{H}_{\text{S}}$ has been eliminated from
the series in Eq.\eqref{eq:RES05}. Therefore the dimension of the
space spanned by the bound states of the Coulomb system is fewer by this
missing one than that of $\hat{H}_{\text{S}}$ with $E=-\omega^{2}/2$.
Since $\hat{H}_{\text{S}}$ is Hermitian, its eigenvectors form a
complete set. Hence the bound states of $\hat{H}_{\text{C}}$ cannot be
complete. This will become clear if we set $\alpha=0$, by which we
return to the free particle, in Eq.\eqref{eq:RES17}. The sum over
discrete eigenvalues is missing a term of $n=0$, thereby failing to
reproduce a delta function $\delta(p_{b}-p_{a})$. Interestingly, the
missing term is supplied from the term of continuous eigenvalues. The
same can be seen more clearly in the integral of Eq.\eqref{eq:RES23} for
the resolvent operator of a positive energy. For this case, if
$\alpha\ne0$, poles from $\nu/(e^{2\pi\nu}-1)$ for bound states are
complemented by the contribution from the pole of $1/(\nu-\hat{\alpha})$
for continuous eigenvalues to form a complete set. If we set $\alpha=0$
after shifting $\nu$ by $-i/2$($+i/2$) for the first(second) term,
the pole at $\nu=\hat{\alpha}$ is absorbed into those of
$1/(1+e^{2\pi\nu})$ by adding new poles at $\nu=\pm i/2$. Cauchy's
integral theorem then tells us that the sum over these poles at
$\nu=\pm i(n+1/2)$($n=0,\,1,\,2,\,\dots$), with suitable regulators for 
each sum from upper and lower half-planes, is entirely equivalent to the
integration of the continuous eigenfunctions over $\nu$ along the real
axis, therefore resulting in a delta function $\delta(p_{b}-p_{a})$ from
both approach. It is exactly the pole at $\nu=\hat{\alpha}$ in
Eq.\eqref{eq:RES23} that generates the term of continuous eigenvalues in 
Eq.\eqref{eq:RES17}. Therefore, we are convinced the incompleteness of
the set of eigenfunctions of bound states of the Coulomb system. It will
be further interesting to notice that the eigenvalue, $E_{0}=-\infty$,
of the missing discrete eigenfunction is connected to that of
$E=k^{2}/2$ by letting $k\to\infty$ because, if $\alpha\ne0$,
$\hat{\alpha}=0$ is possible only in this limit. Namely, the scattering
state with an energy of positive infinity is complementary to the bound
state with an energy with negative infinity though unphysical they are.
A very intuitive understanding for this will be possible if we remember
the behaviors of eigenfunctions in the coordinate representation; both
limits, $n\to0$ so that $\omega_{n}\to-\infty$ for the bound state and
$k\to\infty$ for the scattering state, yield a delta function
$\delta(x)$ although approaching in different directions. Finally, the
absence of the bound state for $n=0$ in Eq.\eqref{eq:RES17} may be
regarded as the consequence of the boundary condition imposed naturally
by the formulation of the path integral for wave functions at $x=0$. In
view of the behavior of eigenfunctions at $x=0$, we recognize that the
path integral requires the Dirichlet condition at the origin.

\section{Conclusions}
On the basis of the Duru-Kleinert transformation with Fujikawa's gauge
theoretical technique, we have formulated exact path integrals
for the Coulomb potential in one space dimension. The mechanism of the
exactness is explained by the pole structure of kernels in the
momentum representation so that the time-sliced path integrals are
evaluated simply by the use of Cauchy's integral theorem. Therefore the
model considered here will be regarded as a new example of the path
integral that can be performed exactly without making use of the
Gaussian integration. It will be, however, difficult to observe such a
structure if we employ the formal continuum formulation path
integrals. A special care on the bound state with infinite binding
energy has been taken to clarify its absence. The disappearance of this
unphysical state in the eigenfunction expansion of the resolvent
operator should be attributed to a proper prescription for the boundary
condition at the origin required by the path integral formalism.

The mechanism of the exact path integrals may have another
interpretation; each relation between $p_{a}$ and $p_{b}$ read from the
poles of the kernels in the momentum representation defines an
action of the one parameter subgroup of $SL(2,\bm{R})$(setting $\tau=it$
is assumed for the case of a negative energy) through a linear
fractional transformation. 
This will bring us a group theoretical approach for the understanding
of the exactness. From this view point, generalizations of our technique
to be useful for other systems will be interesting for future
investigation. 

\section*{Acknowledgments}

The author would like to thank his students, S. Nishioka and
T. Yamashita, for fruitful discussions. He also wishes to express his
gratitude to Professor R. Loudon of the University of Essex for a
kind comment\cite{Loudon08} on the physical reasons for his own approach
to the problem, grounded in measurements of exciton
spectra\cite{ElliottLoudon} made long ago.


\end{document}